\documentclass[pre,twocolumn,superscriptaddress]{revtex4}

\usepackage{anysize}
\marginsize{2cm}{2cm}{1.cm}{1.cm}
\usepackage{graphicx}
\usepackage{epsfig}
\usepackage{verbatim}
\usepackage{amssymb}
\usepackage{amsfonts,amsmath}
\usepackage{bm}
\usepackage[T1]{fontenc}
\usepackage{eufrak}
\usepackage[colorlinks]{hyperref}
\usepackage{fancyhdr}
 \usepackage{relsize}

\hypersetup{colorlinks=true,
	linkcolor=black,
	anchorcolor=black,
	citecolor=black,
	urlcolor=black
}

\providecommand{\bw}{\begin{widetext}}
\providecommand{\ew}{\end{widetext}}
\providecommand{\bal}{\begin{aligned}}
\providecommand{\eal}{\end{aligned}}
\providecommand{\brho}{{\bm \rho}}

\newcommand{\nn}{\nonumber}

\newlength{\bilderlength}

\newlength{\figsize}
\newcommand{\st}{\scriptscriptstyle}

\begin{document}

\author{Jacky Nguyen}
\affiliation{Laboratoire Physico-Chimie Curie, Institut Curie, CNRS UMR168, 75005 Paris, France}
\author{Michele Castellana}
\affiliation{Laboratoire Physico-Chimie Curie, Institut Curie, CNRS UMR168, 75005 Paris, France}

\setlength{\parskip}{5pt plus 0pt minus 0pt}
\title{Optimal localization patterns in bacterial protein synthesis}
\begin{abstract}
In \textit{Escherichia coli}  bacterium, the molecular compounds involved in protein synthesis, messenger RNAs (mRNAs) and ribosomes, show marked intracellular localization patterns. Yet a quantitative understanding of the physical principles which would allow one to control protein synthesis by designing, bioengineering, and optimizing these localization patterns is still lacking. 
In this study, we consider a scenario where a synthetic modification of mRNA reaction-diffusion properties allows for controlling the localization and stoichiometry of mRNAs and polysomes---complexes of multiple ribosomes bound to  mRNAs. Our analysis demonstrates that protein synthesis  can be controlled, e.g., optimally enhanced or inhibited, by leveraging mRNA spatial localization and stoichiometry only, without resorting to alterations of mRNA expression levels. We identify the  physical mechanisms that control the protein-synthesis rate, highlighting the importance of  colocalization between mRNAs and freely diffusing ribosomes, and the interplay between polysome stoichiometry and excluded-volume effects due to the DNA nucleoid. 
The genome-wide, quantitative predictions of our work may allow for a direct verification and implementation in cell-biology experiments, where localization patterns and protein-synthesis rates may be monitored by fluorescence microscopy in single cells and populations. 
\end{abstract}

\maketitle

\section{Introduction}\label{intro}

Bacterial cells exhibit a marked internal organization: Rather than uniformly filling the entire cell volume, a variety of biomolecules display distinct, specific intracellular localization patterns \cite{gahlmann2014exploring}. Notable examples are chromosome partitioning in \textit{Caulobacter crescentus} \cite{ptacin2010spindle}, as well as membrane localization of permease proteins \cite{nevo-dinur2011translation} and localization of cell-division proteins in a constriction ring at midcell \cite{fu2010vivo} in \textit{Escherichia coli} (\textit{E. coli}) bacterium. 

In recent years, an increasing number of studies focused on the localization patterns of  the \textit{E. coli} translational machinery \cite{golding2004rna,mondal2011entropy,bakshi2012superresolution,bakshi2014time,sanamrad2014single}, which is composed of polymeric molecules---messenger RNAs (mRNAs)---and ribosomes---molecular complexes that bind to and translate mRNAs, and thus synthesize proteins. These experimental developments call for a theoretical understanding of the physical mechanisms  which govern such spatial localization. In particular, despite the recent insights above, we still lack  genome-wide, quantitative insights into the physical principles which would allow one to design, bioengineer and optimize  such spatial-localization patterns, in an effort to control and regulate protein synthesis. 

In thus study, we address this matter by introducing a reaction-diffusion model for the concentrations of free ribosomes, mRNAs, and polysomes---complexes of multiple ribosomes bound to  mRNAs. We show that the protein-synthesis rate can be controlled by leveraging mRNA spatial localization  and polysome stoichiometry, without resorting to alterations of mRNA expression levels. The resulting optimal profiles identify the fundamental physical features which either enhance or inhibit protein synthesis, highlighting the importance of colocalization between mRNAs and free ribosomes, and of polysome stoichiometry. 

The rest of the paper is organized as follows: In Section \ref{sec2} we introduce the reaction-diffusion model for the ribosome and mRNA dynamics. In Section \ref{sec3} we discuss the  physical constraints to which mRNA concentrations are subjected, define the  protein-synthesis rate as a functional of these concentrations, and thus present the rate optimization in terms of a constrained optimization problem, whose results are presented in Sec. \ref{sec5}. Finally, Section \ref{disc} is devoted to the discussion and physical interpretation of the optimal mRNA configurations, to the resulting quantitative gain for the protein-synthesis rate, and to an outlook on the  realizations of such configurations in bioengineering experiments.

\begin{figure}
\centering\includegraphics[scale=.8,angle=00]{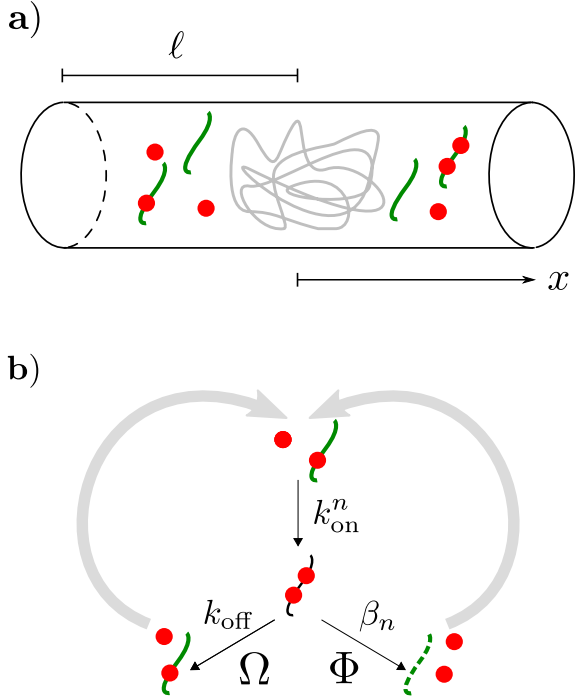}
\caption{
Illustration of the \textit{Escherichia coli} (\textit{E. coli}) geometry and translational machinery.  a) We model an \textit{E. coli} cell as  a cylinder of half length $\ell$ and radius $R$, where the coordinate $x$ runs along the long cell axis. Ribosomes are represented by red spheres, messenger RNAs (mRNAs) by green solid curves,   polysomes by ribosome-mRNA complexes, and the gray closed curve at midcell denotes the DNA nucleoid. b) Schematic of the ribosome reaction process. Ribosomes bind to  and translate mRNAs at a rate $k_{\rm on}^n$ (top arrow) and  form polysomes with $n$ bound ribosomes.  A ribosome on a polysome can either complete mRNA translation and unbind from the mRNA at a rate $k_{\rm off}$ (left arrow), or be freed before translation completion because the mRNA  is degraded (green dashed curve), where $\beta_n$ is the degradation rate (right arrow). 
The total rate at which ribosomes complete translation in the right half of the cell is denoted by $\Omega$, while $\Phi$ is the total rate in the right half of the cell at which ribosomes are freed from mRNAs because of mRNA degradation. 
Given that the total number of ribosomes is conserved, freed ribosomes may rebind  to mRNAs (gray wide arrows).
 \label{fig3}}
\end{figure}

\section{Reaction-diffusion model}\label{sec2}

We model an \textit{E. coli} cell as a cylinder with half length $\ell$ and radius $R$, and introduce the coordinate $x$ which runs along the long cell axis, where $x=0$ corresponds to the cell center and $x=\ell$ to the right pole of the cell \cite{castellana2016spatial},  see Fig. \ref{fig3}a. 
Given the observed symmetry between  left and right half of the cell \cite{castellana2016spatial}, we focus on the region $0 \leq x \leq \ell$, and introduce the  one-dimensional concentration of free (F) ribosomes  $c_{\rm \st F}(x)$, where $c_{\rm \st F}(x) dx$ is the number of ribosomes in a cell slice between $x$ and $x+dx$. 
The reaction-diffusion equation for the concentration of F ribosomes reads \cite{castellana2016spatial}
\begin{equation}
\bal\label{eqF}
\frac{\partial c_{\st \rm F}(x,t)}{\partial t} &=&   \\ 
D_{\st \rm F} \left[ \frac{\partial^2 c_{\st \rm F}(x,t)}{\partial x^2} v_{\st \rm F}(x) -  c_{\st \rm F}(x,t) \frac{d^2 v_{\st \rm F}(x)}{d x^2} \right]  && \\ 
  -  c_{\st \rm F}(x,t) \hspace{-1mm}   \sum_{n=0}^{n_{\rm max}-1}k_{\rm on}^n \rho_n(x,t)  \hspace{-0.5mm}+\hspace{-0.5mm} k_{\rm off} \hspace{-1mm}  \sum_{n=1}^{n_{\rm max}}    n  \, \rho_{n}(x,t) && \\ 
 +   \sum_{n=1}^{n_{\rm max}}  \beta_n \,n \,   \rho_n(x,t), &&
\eal
\end{equation}
where the term in the second line represents diffusion in the presence of excluded-volume effects due to the DNA nucleoid, and $D_{\rm \st F}$ is the diffusion coefficient.
We observe that Eq. (\ref{eqF}) describes the dynamics of F ribosomes in the presence of  a given, time-independent DNA concentration profile \cite{castellana2016spatial}: as a result $v_{\rm \st F}(x)$, i.e., the fraction of available volume within the DNA mesh at position $x$, is also fixed, see  Appendix \ref{app1} for details. 

In addition to diffusion, Eq. (\ref{eqF}) includes ribosome-mRNA reaction terms: The first term in the third line represents ribosome-mRNA binding. This quantity is proportional to the  rate $k_{\rm on}^n$ at which F ribosomes bind to polysomes composed of mRNAs  and $n$ bound ribosomes,  whose density is denoted by $\rho_n$. In addition, $n_{\rm max}$ denotes the maximum number of ribosomes that can be loaded on an mRNA. The second term in the third line represents unbinding of ribosomes due to translation completion, where the unbinding rate is $k_{\rm off}$, and the factor $n$ accounts for the fact that unbinding from a polysome with $n$  ribosomes can happen in $n$ ways. Finally, the last line describes mRNA degradation, where $\beta_n$ is the degradation rate of polysomes with $n$  ribosomes.

We introduce the F-ribosome flux 
\begin{equation}\label{eq105}
J(x,t) = -D_{\st \rm F} \left[\frac{\partial c_{\st \rm  F}} { \partial x}  v_{\st \rm F}(x) - c_{\st \rm F}(x,t) \frac{d v_{\st \rm F}}{dx } \right],
\end{equation} 
and  combine Eq. (\ref{eqF}) with the boundary condition of zero flux at the right cell pole

\begin{equation}\label{eq101}
J(\ell,t)=0,
\end{equation}
and with a constraint which enforces the conservation of the total number of ribosomes \cite{castellana2016spatial}
\begin{equation}\label{eq102}
2\int_0^\ell dx \left[c_{\st \rm F}(x,t) + \sum_{n=1}^{n_{\rm max}} n \rho_n(x,t)\right] = N_{\rm tot}.
\end{equation} 
Overall, the ribosome reaction processes in Eq. (\ref{eqF}) and the conservation of the total number of ribosomes can be schematically represented by the diagram in Fig. \ref{fig3}b.

In order to introduce the protein-synthesis optimization, it  is useful to recall the reaction-diffusion equations which describe the dynamics for the mRNA concentrations \cite{castellana2016spatial}:
\bw
\begin{equation}
\bal \label{eqs1}
\frac{\partial \rho_n (x,t)}{\partial t} =&  D \left[ \frac{\partial^2 \rho_n (x,t)  }{\partial x^2}  v_n(x) - \rho_n(x,t) \frac{d^2 v_n(x)}{d x^2} \right] \\ 
 & - \mathbb{I}(n<n_{\rm max}) k^n_{\rm on} c_{\st \rm F}(x,t) \rho_n(x,t)  -  k_{\rm off} \, n \,  \rho_n(x,t)   +  \mathbb{I}(n>0) k^{n-1}_{\rm on} c_{\st \rm F}(x,t) \rho_{n-1}(x,t)  \\ 
& + \mathbb{I}(n<n_{\rm max})\, k_{\rm off} \, (n+1)  \,  \rho_{n+1}(x,t)  + \mathbb{I}(n=0) \, \alpha(x) - \beta_n \, \rho_n(x,t)
\eal
\end{equation}
\ew
with no-flux boundary conditions 
\begin{equation}\label{eq100}
-D \left[\frac{\partial \rho_n} { \partial x}  v_n(x) - \rho_n(x,t) \frac{d v_n}{dx } \right] =0
\end{equation}
at $x=0$ and $x = \ell$ \cite{castellana2016spatial}. The term in the right-hand side in the first line of Eq. (\ref{eqs1}) represents diffusion,  $D$ is the diffusion coefficient, and $v_n$ is the available volume for polysomes with $n$ ribosomes, see Appendix \ref{app1} for details. The terms in the second line and the first term in the third  line describe ribosome binding and unbinding, and they are analogous to the reaction terms in Eq. (\ref{eqF}). The remaining terms describe synthesis of ribosome-free mRNAs with rate $\alpha(x)${, which is discussed in Appendix \ref{app4}, }and mRNA degradation with rate $\beta_n$. Finally,   the indicator function $\mathbb I$ is equal to unity when the condition in its argument is satisfied and zero otherwise, and it singles out  the terms compatible with the condition $0 \leq n \leq n_{\rm max}$.

\section{Optimization of protein-synthesis rate}\label{sec3}

We now focus on the steady state,  set
\[
\frac{\partial c_{\st \rm F}(x,t)}{\partial t} = \frac{\partial \rho_n (x,t)}{\partial t}=0,
\]
and thus drop the time dependence in all quantities. At steady state, Eqs.  (\ref{eqF}) and (\ref{eqs1}) determine the concentrations of  F ribosomes and mRNAs. In particular,   the steady-state mRNA profiles are set by the reaction-diffusion, creation and degradation processes   to which mRNAs are subject, see the right-hand side of Eq. (\ref{eqs1}).  We observe that a synthetic alteration of such  processes, e.g.,  a change of the diffusion coefficient or the inclusion of a reaction term reflecting the interaction with additional molecular species, would allow one to alter the mRNA dynamics, and thus the mRNA concentrations. One could thus imagine to induce a synthetic, steady-state mRNA profile, and study the  behavior of F ribosomes in the presence of these profiles by means of Eq. (\ref{eqF}), which yields $c_{\rm \st F}$ as a function of the mRNA densities 
 \[
 \brho(x) \equiv (\rho_0(x), \cdots, \rho_{n_{\rm max}}(x)).
 \]
 In particular,  we will study how the protein-synthesis rate  may be controlled, i.e., maximized or minimized, by altering $\brho$. 

To achieve this, we impose a set of minimal, physical conditions on the synthetic mRNA profiles above, which need to be satisfied for such profiles to be realizable.

First, at steady state the overall rate of mRNA synthesis and degradation must be equal
\begin{equation}\label{eq5}
\alpha_{\rm tot} = \int_0^\ell dx \sum_{n=0}^{n_{\rm max}} \beta_n \rho_n(x),
\end{equation}
where
\begin{equation}\label{eq206}
\alpha_{\rm tot} \equiv \int_0^\ell dx \, \alpha(x). 
\end{equation}
Importantly, Eq. (\ref{eq5}) may  be  regarded as a constraint on the total number of mRNAs, which reflects the balance between mRNA synthesis and decay at steady state. 

Second, the overall rate at which ribosomes bind to mRNAs must be equal to the  rate at which ribosomes are freed by either  mRNA unbinding, or by degradation of the mRNA molecule.  This condition follows from the mRNA reaction-diffusion equations, and it can be obtained by integrating Eqs. (\ref{eqs1}) at steady state with respect to $x$, using the boundary conditions (\ref{eq100}), multiplying both sides of Eqs. (\ref{eqs1})  by the ribosome loading number $n$, and summing with respect to $0\leq n \leq n_{\rm max}$. As a result, we obtain the condition
\begin{eqnarray}\nn \label{eq2}
 \int_0^{\ell} dx \, c_{\st \rm F}(x)  \sum_{n=0}^{n_{\rm max}-1}k_{\rm on}^n \rho_n(x) & =& \\ 
  \int_0^\ell dx\,  \sum_{n=1}^{n_{\rm max}} ( k_{\rm off}   +  \beta_n ) n  \, \rho_{n}(x)&,& 
\end{eqnarray}
where the left-hand side denotes the binding rate, and the first and second term in the right-hand side represent the rate at which ribosomes are freed from mRNAs by translation completion and mRNA degradation, respectively. In addition, the rate-balance condition (\ref{eq2}) ensures that the concentration profile of F ribosomes is well behaved at midcell, i.e., that there is a zero F-ribosome flux at $x=0$, see Appendix \ref{app3} for details.

A further constraint results from the finite size of both F ribosomes and polysomes, which yields an upper bound for the total volume that they occupy:
\begin{equation}\label{eq3}
c_{\rm \st F}(x) \frac{4 \pi }{3}r^3+\sum_{n=0}^{n_{\rm max}} \rho_n(x) \frac{4 \pi }{3}r_n^3  \leq \pi R^2
\end{equation}
for $0 \leq x \leq \ell$, where $r$ and $r_n$   are the radii of F ribosomes and polysomes with $n$ ribosomes, respectively, see Appendix \ref{app1}.  

Finally, we consider the excluded-volume effects due to the DNA nucleoid, which prevent large molecules from fitting  in the DNA-rich region. In a cell slice between $x$ and $x + dx$, the total volume occupied by polysomes of species $n$ is $\rho_n(x) dx \frac{4 \pi }{3}r_n^3$. Given that $v_n(x)$ is the probability that a polysome fits in the DNA mesh, the  volume available to polysomes of species $n$ may be estimated as $\pi R^2 dx\,  v_n(x)$. The condition that the occupied volume cannot exceed the available volume thus leads to the constraint
\begin{equation}\label{eq1}
\rho_n(x)  \frac{4 \pi }{3}r_n^3 \leq \pi R^2 v_n(x), 
\end{equation}
which holds for $ 0 \leq n \leq n_{\rm max}$ and $0 \leq x \leq \ell$. \\

We now consider the quantity 
\begin{equation}\label{eq6}
\Omega[\brho] \equiv k_{\rm off} \int_0^\ell  dx \sum_{n=1}^{n_{\rm max}} n \rho_n(x) 
\end{equation}
which, according to Eq. (\ref{eqF}), is the total rate, in the right half of the cell, at which translating ribosomes complete mRNA translation. 
We thus choose $\Omega$ as a proxy for the overall protein-synthesis rate, and seek for the mRNA profiles $\brho$ that either maximize or minimize  $\Omega[\brho]$, and which satisfy  conditions (\ref{eq5}), (\ref{eq2}), (\ref{eq3}) and (\ref{eq1}). 
From the mathematical standpoint, such constrained maximization or minimization constitutes a functional optimization problem with  equality and inequality constraints, which can be studied with known  methods \cite{karush1939minima,kuhn1951nonlinear},  see Section \ref{sec5} and Appendix \ref{app2} for details.

The model parameters have been obtained from experimental data as follows. We consider \textit{E. coli} cells in the midphase of the division cycle with a typical length of $2 \ell = 3 \, \mu \rm m$ \cite{sanamrad2014single},   $N_{\rm tot} = 6 \times 10^4$ ribosomes  { and $N_{\rm mRNA} = 5 \times 10^3$ mRNAs }\cite{castellana2016spatial}. Single F-ribosome tracking  provides the estimate  $D_{\rm \st F} = 0.4\,  \mu \rm m ^2 / \rm s$ for the F-ribosome diffusion coefficient   \cite{sanamrad2014single}, and the available-volume profiles $v_{\rm \st F}(x)$ have been estimated from data on intracellular DNA density profile---see Appendix \ref{app1}.  The  rate at which F ribosomes bind to polysomes of species $n$ is taken to be 
\begin{equation} \label{eq202}
k^n_{\rm on}  = k^{\ast}_{\rm on} \left(1 - \frac{n}{N}\right),
\end{equation}
where $k^{\ast}_{\rm on} \approx 6 \times 10^{-4} \mu \rm m / \rm s$ is the average binding rate across all polysome species, $N = 100$ is an estimate of the maximum number of ribosomes that can be loaded on a typical mRNA \cite{castellana2016spatial}, and the factor $1 - n /N$ estimates the decrease of the ribosome binding probability with respect to polysome occupancy \cite{zouridis2007model}. The unbinding rate $k_{\rm off} \approx 2.5 \times 10^{-2} / \rm s$ has been estimated as the inverse of the average translation-elongation time \cite{siwiak2013transimulation}.  The mRNA degradation rate 
\begin{equation}\label{eq203}
\beta_n =\beta_{\ast} \left(1 - \frac{n}{N}\right)   
\end{equation}
 has been estimated in terms of the average degradation rate $\beta_{\ast} \approx 3 \times 10^{-3} / \rm s$, which corresponds to a typical mRNA lifetime of  $\sim 5\, \rm min$ \cite{bernstein2004global}. The  factor $1-n/N$ estimates the dependence of the degradation rate on the loading number $n$ in terms of the binding probability of RNAse---the enzyme that initiates mRNA degradation  \cite{mackie2013rnase}. In fact, mRNA degradation is the result of a competition between RNAse and ribosomes,  which both bind  to the 5' end of the mRNA \cite{chen2015genome}: We  estimate the probability that the 5' site is occupied by a ribosome as the ratio between the ribosome number $n$ and the number $N$ of free sites on the mRNA, and thus estimate the RNAse binding probability as $1-n/N$. 

\section{Results}\label{sec5}

We  estimated numerically the solution of the optimization problem above by discretizing the long cell axis in $L$ bins $x_1, \cdots, x_L$, and representing all concentrations, e.g., ${\bm \rho}(x)$, as  $L (n_{\rm max}+1)$-dimensional vectors ${\bm \rho}(x_1), \cdots, {\bm \rho}(x_L)$. The rate $\Omega[\brho]$ and the constraints then become multivariate functions of  $\{ {\bm \rho}(x_i) \}$, and Eqs.  (\ref{eq5}), (\ref{eq2}), (\ref{eq3}),  (\ref{eq1}) and (\ref{eq6})  are converted into a multidimensional constrained optimization problem, whose solution can be estimated numerically with known  methods, see Appendix \ref{app2} for details.  In what follows, we will show the optimization results for both the rate maximization and minimization, whose physical interpretation will be discussed in Section \ref{disc}. 

\begin{figure*}
\centering\includegraphics[scale=1.12]{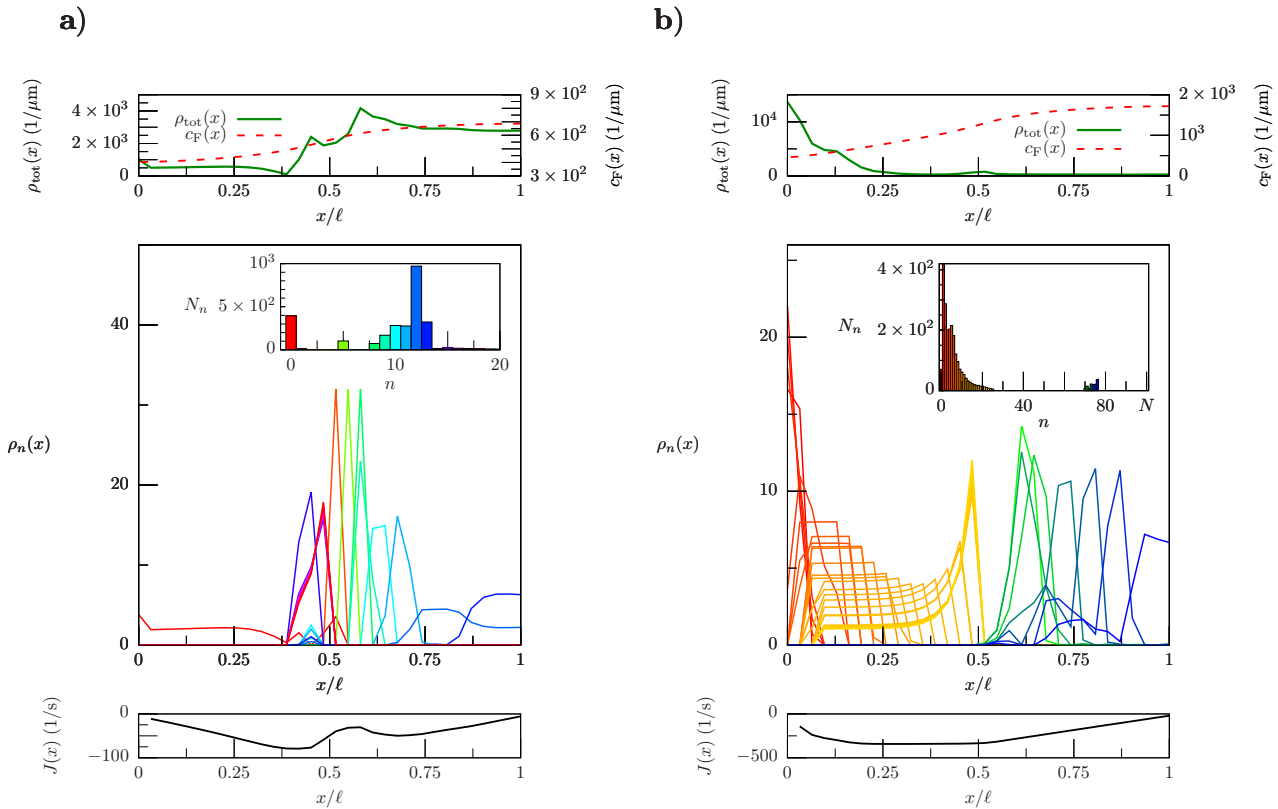}
\caption{
{ Numerical estimate of the optimal profiles, total number of mRNAs and fluxes for the protein-synthesis rate maximization and minimization. } a) Results for the rate maximization. Top: Total mRNA density $\rho_{\rm tot}(x)$ (green solid curve) and  free-ribosome concentration  $c_{\rm \st F}(x)$ (red dashed curve)  as functions of the ratio between the coordinate $x$ on the long cell axis and $\ell$, i.e., half the cell length, in the right half of the cell, see Fig. \ref{fig3}a.  
Middle: mRNA densities $\rho_n(x)$  as functions of $x/\ell$. The densities $\rho_n$ are normalized to unit area and are thus dimensionless, and their colors correspond to those of the mRNA species  in the inset, where we show the total mRNA number $N_n = \int _0 ^\ell d x \rho_n(x)$  in the right half of the cell as a function of $n$. Profiles $\rho_n$ such that $N_n$ is small enough not to be visible in the inset are not shown for clarity. 
Bottom: free-ribosome flux as a function of $x/\ell$. b)  Results for the rate minimization,  see a) for details.  The maximal number $N$ of ribosomes that can be loaded on an mRNA  is  marked in the inset. 
\label{fig2}}
\end{figure*}

The results of the rate maximization are shown in  Fig. \ref{fig2}a, for a maximal ribosome number $n_{\rm max} = 48$ and a bin number $L = 32$, see Fig. \ref{figs1}a for a   convergence check of the results with respect to $L$. 
The optimal protein-synthesis rate is $\Omega_{\rm \st MAX} = 728.79 / \rm s$. The top panel  shows a marked, overall colocalization of mRNAs and F ribosomes: the total mRNA density $\rho_{\rm tot}(x) = \sum_{n=0}^{n_{\rm max}} \rho_n(x)$  and F-ribosome profile $c_{\rm \st F}(x)$ are both concentrated at the cell poles. In the middle panel, we show the polysome densities $\rho_n(x)$ vs. $x/\ell$ and the total number of polysomes of species $n$ in the right half of the cell  
\begin{equation}\label{eq205}
N_n \equiv \int_0^ \ell dx \rho_n(x)
\end{equation}
as a function of $n$. The polysome number $N_n$ is peaked around two values: Ribosome-free mRNAs with $n=0$, which are localized within the nucleoid, and polysomes loaded with $n \sim 12$ ribosomes, which are localized at the  poles. The matching colors in  the top panel and inset show that the larger the ribosome loading number $n$, the more  $\rho_n(x)$ is localized outside the nucleoid---a feature which follows from the excluded-volume constraint (\ref{eq1}). The bottom panel shows that the F-ribosome flux $J(x)$ is negative throughout the entire right half of the cell, i.e., F ribosomes flow from the cell poles to midcell. 

Figure \ref{fig2}b shows the optimization results for the rate minimization, with $n_{\rm max} = 100$ and $L=32$, see Fig. \ref{figs1}b for a convergence check of the results with respect to $L$. The optimal protein-synthesis rate is $\Omega_{\rm \st MIN} = 704.96 / \rm s$, which is $\sim 3\%$ smaller than the maximal rate $\Omega_{\rm \st MAX}$.  Unlike the maximization  above, the top panel shows that mRNAs are antilocalized from F ribosomes: The total mRNA profile is mostly concentrated at midcell, while  F ribosomes are localized at the poles. In the middle panel, the main plot shows that the larger the ribosome  number $n$, the more the density $\rho_n(x)$ is localized at the cell poles, see Fig. \ref{fig2}a. The polysome distribution in the inset is mostly concentrated around $n=0$, with an additional peak at $n \sim 75$, and  the bottom panel shows that the F-ribosome flux is directed towards the cell center.

\section{Discussion}\label{disc}

The recent interest in intracellular localization of messenger RNAs (mRNAs)  in both eukaryotic \cite{herve2004zipcodes} and prokaryotic \cite{golding2004rna,mondal2011entropy,pilhofer2009fluorescence,buxbaum2014right}   organisms has led to growing investigations on the connection between mRNA localization and  cellular functions. These include sorting of different protein products to specific subcellular locations \cite{kannaiah2014protein}, such as co-translational membrane insertion of nascent proteins by mRNAs localized near the membrane \cite{bakshi2012superresolution,korkmazhan2017dynamics}. 

In this study, we analyze the impact on the protein-synthesis rate of spatial localization and stoichiometry of mRNAs in \textit{Escherichia coli} (\textit{E. coli}) bacterium. We consider the reaction-diffusion dynamics for free (F) ribosomes and mRNAs, written in terms of the reaction-diffusion equations for their concentrations  in the presence of a given DNA concentration profile. Such dynamics set the localization profiles of F ribosomes, mRNAs and polysomes, i.e., mRNAs loaded with multiple bound ribosomes, and  determine the polysome stoichiometry---the distribution of polysomes with different loading numbers \cite{castellana2016spatial}.  We investigate a scenario where a synthetic modification of mRNA reaction-diffusion properties allows for controlling the localization of mRNAs and polysomes and their stoichiometry. Free ribosomes would then obey a reaction-diffusion dynamics in the presence of a given configuration of  mRNA and polysome concentration profiles, and the protein-synthesis rate would be a functional of such concentrations. 
We then seek the mRNA and polysome profiles that either maximize or minimize such synthesis rate, and which satisfy a set of fundamental physical conditions. Such conditions imply, for example, a constraint on the total number of mRNAs, and thus on their expression levels.

The results of this functional, constrained optimization problem in Fig. \ref{fig2}a show that the profiles which maximize the synthesis rate bear a striking resemblance to those resulting from the full, reaction-diffusion dynamics of F ribosomes, mRNAs and polysomes, which supposedly represent the actual profiles in live \textit{E. coli} cells \cite{castellana2016spatial,golding2004rna}. In this optimal solution, the synthesis rate  is enhanced by colocalizing F ribosomes and heavily loaded polysomes at the cell poles.

In addition, the  optimal polysome distribution in Fig. \ref{fig2}a may be interpreted as follows. As shown in Appendix \ref{app4}, the constraints of the optimization problem  imply that the following combination of   protein-synthesis and mRNA-degradation rates
\begin{equation}\label{eq207}
\left( 1+\frac{ k_{\rm on}^\ast \alpha_{\rm tot}}{ \ell \beta_\ast k_{\rm off}} \right) \Omega + \Phi
\end{equation}
is approximately constant throughout the optimization, where 
\begin{equation}\label{eq209}
\Phi[{\bm \rho}] \equiv \sum_{n=1}^{n_{\rm max}}\beta_n n N_n
\end{equation}
denotes the total mRNA degradation rate, see Fig \ref{fig3}b. It follows that the protein-synthesis rate mat be enhanced by either increasing $\Omega$ itself (I), or by decreasing  $\Phi$ (II). These two mechanisms allow us to interpret the polysome distribution in  the inset of Fig. \ref{fig2}a: The peak around heavily loaded polysomes $n\sim 12$ enhances the synthesis rate $\Omega$, and thus reflects the optimization strategy I, while the  peak at $n=0$ hinders the mRNA degradation rate $\Phi$, see mechanism II. \\

Finally,  we observe that, while large polysomes are segregated from midcell due to excluded-volume effects of the DNA nucleoid, the optimization leverages the capability of small, ribosome-free mRNAs to penetrate into the nucleoid by localizing them at midcell. This localization pattern creates an efficient ribosome-recycling process, where ribosomes that unbind from polysomes at the poles flow at midcell and rebind to ribosome-free mRNAs to initiate new translations. \vspace{2mm}\\

{The minimization of the synthesis rate is achieved by means of a segregation of F ribosomes away from mRNAs, which are mostly localized at the poles and midcell, respectively, see  Fig. \ref{fig2}b. 
The polysome stoichiometry in Fig. \ref{fig2}b allows for an interpretation along the lines of the one given for the  rate maximization, where mechanisms I and II now correspond to minimization of $\Omega$ and maximization of $\Phi$, respectively. First, most polysomes are loaded with a small number of ribosomes, thus reducing the protein-synthesis rate according to mechanism I. In addition, such a small loading number allows polysomes  to localize within the nucleoid at midcell, where the excluded-volume effects due to the DNA    prevent F ribosomes from penetrating in the polysome-rich region,  thus lowering further the synthesis rate $\Omega$, in line with mechanism I. Second,   Eq. (\ref{eq209}) shows that the number  $N_n$ of polysomes of species $n$ enters in $\Phi$ with the weight  $n \beta_n$: It follows that $\Phi$ may be maximized by concentrating the polysome distribution at $n = N/2$, where  $n \beta_n \propto n (1-n/N)$ is maximal. As a result, mechanism II would justify the existence of a peak in the polysome distribution at large values of $n \sim N/2$, such as the one located at  $n\approx 70$ in the inset of Fig. \ref{fig2}b. \vspace{3mm}\\}

Finally, we discuss the physical origin for the $\sim 3\%$ difference between the maximal and minimal synthesis rates above. The smallness of this	 difference can be understood by observing that  the time for an F ribosome to diffuse through the entire cell, i.e.,  $(2 \ell)^2 /D_{\rm \st F} \approx 20 \, \rm s$, is significantly shorter than the average mRNA lifetime $ 1/\beta_\ast  \approx 5\,  \rm min$. As a result, localizing mRNAs in a specific intracellular region will have a small effect on the protein-synthesis rate, because ribosome diffusion is fast enough that an F ribosome is likely to reach, bind to and translate the mRNA before this decays, no matter where the mRNA is localized.\\

To summarize, our analysis provides a novel means of controlling protein-synthesis rate in bacterial cell biology which does not resort to alterations of mRNA expression levels,  and leverages uniquely mRNA  localization patterns and  stoichiometry. Our study may provide some guidance for bioengineering efforts on the \textit{E. coli} translational machinery, highlighting the importance of ribosome-mRNA localization for the protein-synthesis rate, and targeting the optimal localization patterns and stoichiometries of mRNAs and polysomes. First, such localization patterns may be engineered by constructing mRNA sequences that code for proteins which bind to given intracellular sites, thus bringing mRNAs close to specific  locations \cite{bakshi2012superresolution}. Second, the insertion of  nonoptimal codons which slow down translation may cause ribosome accumulation on specific mRNA chains, thus increasing their ribosome loading number, and ultimately allowing for a synthetic control of polysome stoichiometry \cite{pechmann2013ribosome}. 
The resulting  mRNA localization patterns  and  protein-synthesis rates may be monitored  and quantified by fluorescence microscopy in single cells and populations \cite{golding2004rna,kourtis2009cell}, thus allowing for a quantitative experimental verification of the model predictions. 
In particular, despite the fact that the relatively small predicted difference in the synthesis rate is likely to be irrelevant in a single cell because of a variety of factors, e.g., thermal noise, fluctuations, and others, this difference may be amplified on a multicellular level. Namely, if the engineered localization pattern in the parent cells could be reproduced in the daughter cells across a sufficiently large number of cell cycles, the effects of localization may be detected on the scale of a bacterial population by measuring, for example, the population size as well as the total protein yield in the population.

\begin{acknowledgments}
We thank  J.-F. Joanny, S. A. Lira, K. Papenfort, S. van Teeffelen and N.  S. Wingreen for useful discussions. The numerical calculations presented in this work were performed on computational resources supported by the Princeton Institute for Computational Science and Engineering (PICSciE), the Office of Information Technology's High Performance Computing Center and Visualization Laboratory, and the Lewis-Sigler Institute for Integrative Genomics at Princeton University. This work was granted access to the HPC resources of MesoPSL financed by the Region \^{l}le de France and the project Equip@Meso (reference ANR-10-EQPX-29-01) of the programme `Investissements d'Avenir' supervised by the Agence Nationale pour la Recherche. 
\end{acknowledgments}

\vspace{1cm}
\appendix

\section{Available volume}\label{app1}

In what follows, we report a summary of the available-volume estimate presented in \cite{castellana2016spatial}, to which we refer for additional details.

Given that F ribosomes have a finite diameter, they are subjected to excluded-volume effects due to the DNA, whose polymeric structure is strongly compacted in the nucleoid at midcell \cite{bakshi2012superresolution}. The probability that a diffusing F ribosome  fits into the DNA mesh at position $x$ is denoted by $v_{\rm \st F}(x)$, and is estimated as follows, see \cite{castellana2016spatial} for details. Based on experimental DNA-fluorescence profiles, we introduce the local density of DNA length 
\begin{equation}\label{eq201}
\varphi(x) \propto \frac{1}{1+\exp[20 (x/\ell-1/2)]},
\end{equation}
where the proportionality constant is determined by the normalization condition $\frac{1}{\ell}\int_0^\ell dx \, \varphi(x) = \frac{L}{V}$, $L = 1.5 \, \rm mm$ is the total DNA plectoneme length \cite{mondal2011entropy}, and $V = 1.2\,  \mu \rm m ^3$ is the volume of the nucleoid \cite{bakshi2012superresolution}. The available volume  reads \cite{castellana2016spatial}
\begin{equation}
v_{\rm \st F}(x) = \exp[ - \pi r^2 \varphi(x)/4 ],
\end{equation}
where  $r \simeq 10 \, \rm nm$  is  the ribosome radius \cite{kaczanowska2007ribosome}. 

Similarly, the available volume for polysomes is
\begin{equation}
v_n(x)  =  \exp[ - \pi r_n^2 \varphi(x)/4],
\end{equation}
where the polysome radius  $r_n$ is estimated with the relation $r_n^2 =  r_{\st \rm R}^2 + n \, r^2$, and $r_{\st \rm R} = 20 \, \rm nm$ is the typical  radius of gyration of an mRNA molecule \cite{gopal2012visualizing}.

\begin{figure*}
\hspace{-.7cm}\centering\includegraphics[scale=5.5]{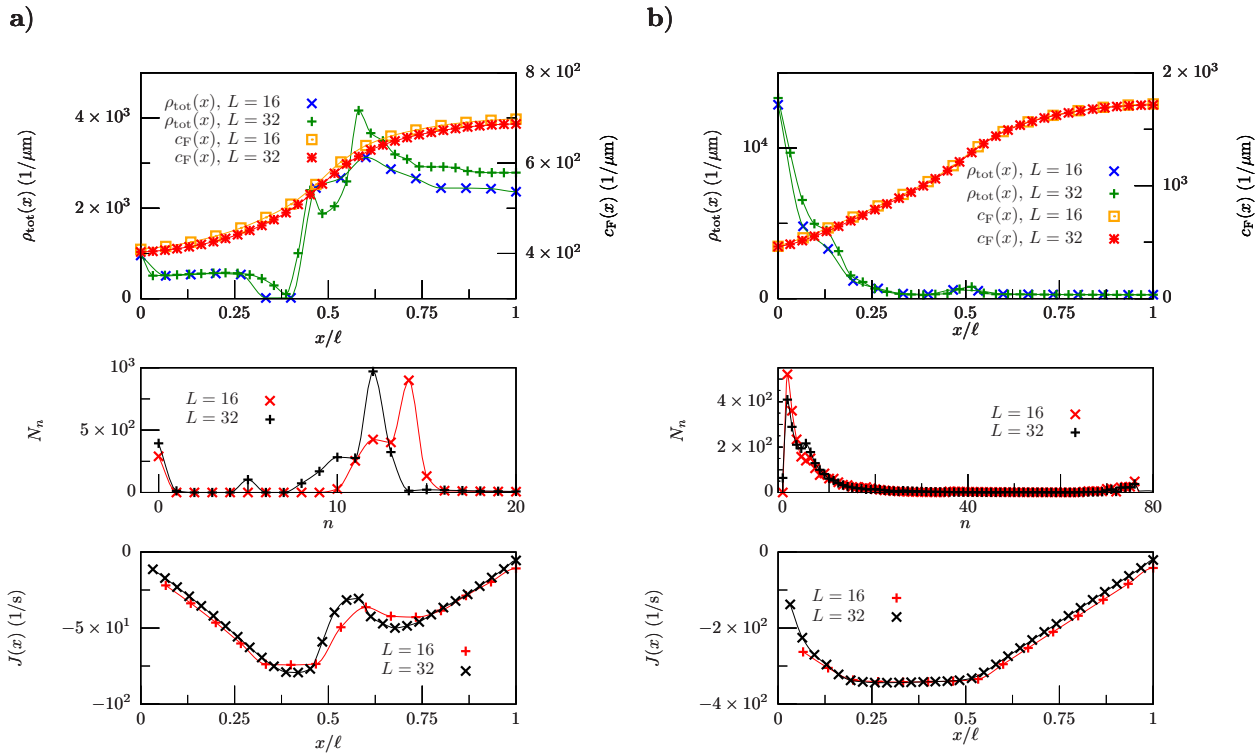}
\caption{
Convergence of the discretization of the reaction-diffusion equations with respect to the number of bins.  a) Total mRNA and free-ribosome densities (top), mRNA number (middle), and free-ribosome flux for the protein-synthesis rate maximization, with $L=16$,  $n_{\rm max}=24$, and with $L=32$, $n_{\rm max}=48$ discretization bins and maximum number of ribosomes per mRNA,  see Fig. \ref{fig2}a for details. b) Same as a) for the minimization of the protein-synthesis rate, where $n_{\rm max} = 100$ for both $L=16$ and $L=32$,  see Fig. \ref{fig2}b. 
\label{figs1}}
\end{figure*}

\section{mRNA synthesis rate}\label{app4}
{
The quantity $\alpha(x)$ describes the  rate at which mRNAs are transcribed at position $x$, and its profile was chosen  to be proportional to the DNA density $\varphi(x)$ in  Eq. (\ref{eq201}). The proportionality constant  was estimated in terms of the total number of mRNAs per cell, $N_{\rm mRNA}$, and of the average mRNA degradation rate $\beta_{\ast}$ introduced in Section \ref{sec3},  by means of the relation $2\int_0^\ell dx \, \alpha(x) / \beta_{\ast} = N_{\rm mRNA}$ \cite{castellana2016spatial}. 
}

\section{Free-ribosome flux at midcell} \label{app3}

In what follows, we will show that the constraint (\ref{eq2}) ensures that there is no F-ribosome flux at midcell, consistently with the left-right symmetry of the cell. By integrating the steady state of Eq. (\ref{eqF}) with respect to $x$ between $x=0$ and $x=\ell$ and using Eq. (\ref{eq105}), we obtain 
\bw
\begin{eqnarray*}\nn 
 J(\ell) - J(0) & = &  \int_0^{\ell} dx \left[- c_{\st \rm F}(x)  \sum_{n=0}^{n_{\rm max}-1}k_{\rm on}^n \rho_n(x) +   \sum_{n=1}^{n_{\rm max}} (  k_{\rm off} + \beta_n) \,n \,   \rho_n(x)\right] \\ \label{eq106}
 & = & -J(0), 
\end{eqnarray*}
\ew
where in the second line we used Eq. (\ref{eq101}). As a result, the constraint (\ref{eq2}) implies that the F-ribosome flux at midcell vanishes, i.e., $J(0) = 0$.

\section{Numerical optimization}\label{app2}
The optimization of the protein-synthesis rate $\Omega[{\bm \rho}]$ was performed with the NLopt package \cite{johnson2010the} by using a gradient-based, sequential least-square quadratic programming algorithm  which allows for a constrained optimization with both equality constraints, Eqs. (\ref{eq5}) and (\ref{eq2}), and inequality constraints, Eqs. (\ref{eq3}) and (\ref{eq1}) \cite{kraft1994algorithm}. 
{ Both equality and inequality constraints allow for a user-defined numerical tolerance, which is  used to assess the constraint fulfillment during the optimization procedure.}
The initial profiles for the optimization were obtained from the steady state of the reaction-diffusion equations (\ref{eqF}), (\ref{eq101}), (\ref{eq102}), (\ref{eqs1}) and (\ref{eq100}). 

\section{Relation between protein-synthesis and mRNA-degradation rate}\label{app4}

\begin{figure*}
\centering\includegraphics[scale=1.7]{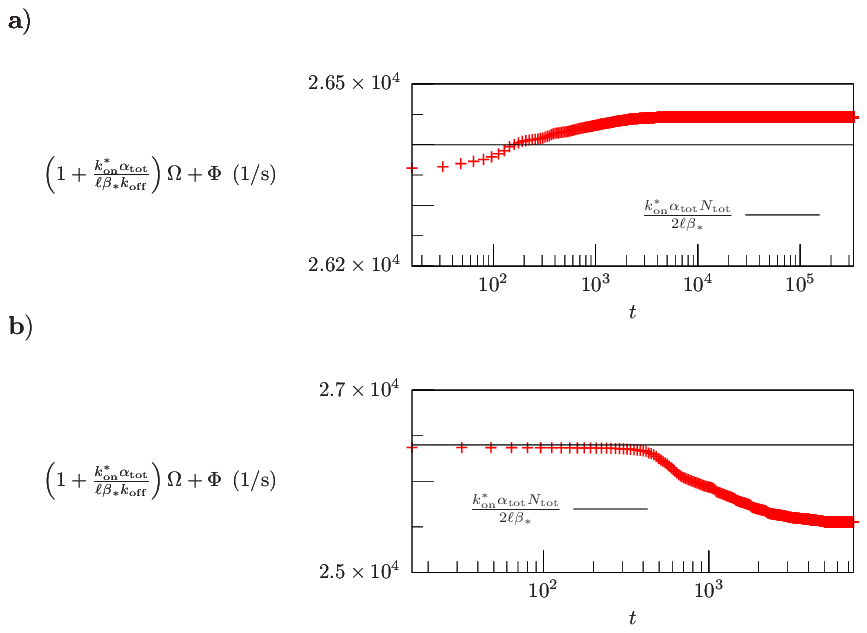}
\caption{\label{figs2}
Illustrative numerical check of Eq. (\ref{eq200}). a) Check for the protein-synthesis maximization, with $L = 16$ and $n_{\rm max} = 24$. The left-hand side of Eq. (\ref{eq200}) (red dots) is shown as a function of the number t of steps in the optimization, together with the right-hand side of Eq. (\ref{eq200}) (black line). b) Same as a) for the minimization, with $L = 16$ and $n_{\rm max}= 100$. 
}
\end{figure*}

In what follows, we will derive an approximate relation between the total rate of protein synthesis and the total mRNA degradation rate. We rewrite constraint (\ref{eq2}) as 
\begin{eqnarray}\label{eq204} \nn
\int_0^\ell dx\,  \sum_{n=1}^{n_{\rm max}} ( k_{\rm off}   +  \beta_n ) n  \, \rho_{n}(x) &\approx &\\ \nn
\frac{N_{\rm \st F}}{2 \ell}  \sum_{n=0}^{n_{\rm max}-1}k_{\rm on}^n  N_n & \approx & \\ \nn
\frac{N_{\rm \st F} k_{\rm on}^\ast}{2 \ell \beta_\ast}  \sum_{n=0}^{n_{\rm max}}\beta_n  N_n & = & \\ \nn
\frac{ k_{\rm on}^\ast \alpha_{\rm tot}}{2 \ell \beta_\ast} N_{\rm \st F} & = &  \\ 
\frac{ k_{\rm on}^\ast \alpha_{\rm tot}}{2 \ell \beta_\ast} \left( N_{\rm tot} - 2 \sum_{n=1}^{n_{\rm max}} n N_n \right)&,& 
\end{eqnarray}
where in the second line we used Eqs. (\ref{eq2}) and  (\ref{eq205}) and observed that, because of the relatively fast diffusion of F ribosomes, the F-ribosome concentration is approximately uniform over the cell: as a result, we replaced $c_{\rm \st F}(x)$ with $N_{\rm \st F}/(2 \ell)$, where $N_{\rm \st F} \equiv 2\int_0^\ell dx c_{\st \rm F}(x)$ is the total number of F ribosomes in the cell. In the third line  we replaced the upper bound of the sum with $n_{\rm max}$, and used Eqs. (\ref{eq202}) and (\ref{eq203}) to rewrite  $k_{\rm on}^n$ in terms of the degradation rate $\beta_n$. In the fourth line we used constraint (\ref{eq5}) and Eq. (\ref{eq206}), and in the fifth line we used the condition (\ref{eq102}) for the total number of ribosomes.  We now combine Eq. (\ref{eq204}) with the definition (\ref{eq6}), and obtain 
\begin{equation}\label{eq200}
\left( 1+\frac{ k_{\rm on}^\ast \alpha_{\rm tot}}{\ell \beta_\ast k_{\rm off}} \right) \Omega +  \sum_{n=0}^{n_{\rm max}}\beta_n n N_n \approx \frac{ k_{\rm on}^\ast \alpha_{\rm tot} N_{\rm tot}}{2 \ell \beta_\ast}
\end{equation}
which, combined with the definition (\ref{eq209}), implies that the quantity in Eq. (\ref{eq207}) is approximately constant throughout the optimization, see Fig. \ref{figs2} for an illustrative numerical check.\\  

\bibliographystyle{phaip}

\end{document}